\pdfoutput=1
\documentclass[prl,twocolumn,showpacs,nofootinbib]{revtex4-1}

\usepackage{graphicx,amsmath,amssymb,amscd,amsfonts,mathtools,moresize}
\usepackage[hidelinks]{hyperref}
\usepackage{xcolor}
\usepackage{ulem} 
\usepackage{thmtools, thm-restate}
\usepackage{epsfig}
\usepackage{epstopdf}
\usepackage{latexsym}
\usepackage{graphicx}
\usepackage{booktabs}
\usepackage{bbm}
\usepackage{color}
\usepackage{physics}
\usepackage{tensor}
\usepackage{verbatim}
\usepackage[caption=false]{subfig}
\usepackage{tikz}
\usepackage{bigints}
\usepackage{ifthen}
\usetikzlibrary{matrix}
\usetikzlibrary{decorations.markings,calc,shapes,decorations.pathmorphing,arrows.meta}
\usetikzlibrary{patterns}
\usetikzlibrary{positioning}
\usepackage{textpos}

\usepackage{hyperref}

\usepackage{xcolor}

\hypersetup{
colorlinks,
linkcolor={blue!80!black},
citecolor={blue!80!black},
urlcolor={blue!80!black},
linktoc=page
}

\begin{document}
\def\nn{\nonumber}
\def\kc#1{\left(#1\right)}
\def\kd#1{\left[#1\right]}
\def\ke#1{\left\{#1\right\}}
\newcommand\beq{\begin{equation}}
\newcommand\eeq{\end{equation}}
\renewcommand{\Re}{\mathop{\mathrm{Re}}}
\renewcommand{\Im}{\mathop{\mathrm{Im}}}
\renewcommand{\b}[1]{\mathbf{#1}}
\renewcommand{\c}[1]{\mathcal{#1}}
\renewcommand{\u}{\uparrow}
\renewcommand{\d}{\downarrow}
\newcommand{\be}{\begin{equation}}
\newcommand{\ee}{\end{equation}}
\newcommand{\bsigma}{\boldsymbol{\sigma}}
\newcommand{\blambda}{\boldsymbol{\lambda}}
\newcommand{\sgn}{\mathop{\mathrm{sgn}}}
\newcommand{\diag}{\mathop{\mathrm{diag}}}
\newcommand{\Pf}{\mathop{\mathrm{Pf}}}
\newcommand{\half}{{\textstyle\frac{1}{2}}}
\newcommand{\sh}{{\textstyle{\frac{1}{2}}}}
\newcommand{\ish}{{\textstyle{\frac{i}{2}}}}
\newcommand{\thf}{{\textstyle{\frac{3}{2}}}}
\newcommand{\SUN}{SU(\mathcal{N})}
\newcommand{\N}{\mathcal{N}}

\renewcommand{\d}{\ensuremath{\operatorname{d}\!}}
\newcommand\dnote[1]{\textcolor{red}{\bf [Daniel:\,#1]}}

\title{Quantum Rods and Clock in a Gravitational Universe}
\preprint{today}
\begin{abstract}
 {Local operators are the basic observables in quantum field theory which encode the physics observed by a local experimentalist. However, when gravity is dynamical, diffeomorphism symmetries are gauged which apparently obstructs a sensible definition of local operators, as different locations in spacetime are connected by these gauged symmetries. This consideration brings in the puzzle of reconciling our empirical world with quantum gravity. Intuitively, this puzzle can be avoided using relatively defined observables when there exists a natural reference system such as a distribution of galaxies in our universe. Nevertheless, this intuition is classical as the rods and clock defined in this way may also have quantum fluctuations so it is not a priori clear if it can be realized in the quantum regime. In this letter, we provide an affirmative answer to this question. Interestingly, we notice that the quantum fluctuations of the reference system are in fact essential for the realization of the above intuition in the quantum regime. }
\end{abstract}
\author{Hao Geng}
\affiliation{Gravity, Spacetime, and Particle Physics (GRASP) Initiative, Harvard University, 17 Oxford Street,
Cambridge MA 02138, USA.}
\maketitle
\section{Introduction}
Locality is one of the basic principles in quantum field theory which reconciles special relativity and quantum mechanics and correctly describes all phenomena we observed so far. In quantum field theory the basic observables are local field operators $\hat{O}(x)$ which are defined at each point $x^{\mu}=(t,\vec{x})$ in the spacetime $\mathcal{M}$. These operators and the smearing of them in a compact spacetime region $\mathcal{U}\subset\mathcal{M}$ encode all the measurements an experimentalist could carry out at the locations and during the moments within $\mathcal{U}$. Nevertheless, when gravity is dynamical the diffeomorphism symmetries
\begin{equation}
    x^{\mu}\rightarrow x'^{\mu}(x)\,,
\end{equation}
are gauged which apparently obstructs the local field operators as gauge invariant observables. More specifically, this obstruction can be seen as follows. Let's consider the infinitesimal version of the diffeomorphism transforms, which in fact generate all local diffeomorphism transforms,
\begin{equation}
    x^{\mu}\rightarrow x^{\mu}+\kappa \epsilon^{\mu}(x)\,,\label{eq:localdiffeo}
\end{equation}
where $\epsilon^{\mu}(x)$ is a local vector field and $\kappa$ is a small parameter indicating the infinitesimal nature of this transform. A scalar field operator $\hat{O}(x)$ transforms under Equ.~(\ref{eq:localdiffeo}) as
\begin{equation}
    \hat{O}(x)\rightarrow \hat{O}(x)+\kappa\epsilon^{\mu}(x)\partial_{\mu}\hat{O}(x)+\mathcal{O}(\kappa^{2})\,,
\end{equation}
which therefore is not invariant as being a local operator its dependence on $x^{\mu}$ is necessarily nontrivial.

The above issue has its root more or less back in the early days of Einstein's theory of general relativity where Einstein visualized the fact that matter curves spacetime by exploiting Mach's Principle. Einstein basically realized that gravitational theory should be a relative theory as a distribution of matter curves spacetime and the curved spacetime tells the observer which frame is inertial, i.e. there is no absolute spacetime.\footnote{In other words, the spacetime diffeomorphism symmetry has to be gauged.} This classical consideration shades much light on its quantum counterpart discussed in the previous paragraph. In simple spacetimes with an asymptotic boundary, it has been shown that we can define quasi-local diffeomorphism invariant observables by dressing them to the asymptotic boundary using the gravitational Wilson lines \cite{Donnelly:2015hta,Donnelly:2016rvo,Donnelly:2017jcd,Donnelly:2018nbv,Giddings:2018umg}. These gravitational Wilson lines are specific line integrals of the graviton field $\hat{h}_{\mu\nu}(x)$ from the spacetime point $x^{\mu}$ to the asymptotic boundary, analogous to Wilson lines in ordinary gauge theories, which have only been constructed for highly symmetric spacetime backgrounds. However, such boundary-dressed operators are only quasi-local as two spacelike separated operators dressed by crossing gravitational Wilson lines don't commute \cite{Donnelly:2015hta}. Furthermore, guided by the intuition that a nontrivial background matter distribution $\phi_{0}(x)$ can be identified as a natural reference frame, it was proposed \cite{Giddings:2005id,Folkestad:2023cze} that a spacetime integral of local operators against a delta-function in the form $\delta(\phi_{0}(x)-a)$, i.e. the delta-function localizes the spacetime integral to the place $x^{\mu}$ where $\phi(x)=a$, should be diffeomorphism invariant. Nevertheless, as an integrated expression, it is not clear if such construction leads to truly localized operators in general backgrounds, as the $\delta$-function might have multiple spacelike separated supports. Therefore, a systematic understanding of whether sensible local operators can be constructed in a gravitational universe is still lacking.

Interesting recent progress in \cite{Geng:2023zhq,Geng:2024tba1,Geng:2024tba2} realizes that in gravitational theories where the graviton is massive due to the spontaneous breaking of diffeomorphism symmetries, one can easily construct diffeomorphism invariant local operators. The construction is done in the fully covariant description and makes use of the Goldstone vector field $\hat{V}^{\mu}(x)$ associated with the spontaneous breaking of the diffeomorphism symmetry. The Goldstone vector field $\hat{V}^{\mu}$, under a proper normalization, transforms under Equ.~(\ref{eq:localdiffeo}) as
\begin{equation}
    \hat{V}^{\mu}(x)\rightarrow \hat{V}^{\mu}(x)+\kappa \epsilon^{\mu}(x)\,,
\end{equation}
so one can visualize $\hat{V}^{\mu}(x)$ as a quantum field of clock and rods and use it to construct diffeomorphism invariant operators. As an explicit example, given a scalar operator $\hat{O}(x)$ we can construct a diffeomorphism invariant operator
\begin{equation}
    \hat{O}^{\text{Phys}}(x)=\hat{O}(x-\hat{V}(x))\,.
\end{equation}
Nonetheless, this construction is rather formal, for example it was not so clear how the vector field $\hat{V}^{\mu}(x)$ emerges in a general scenario with spontaneously broken diffeomorphism symmetries.\footnote{This is understood recently in \cite{Geng:2024tba2} in the so-called \textit{island model}, a setup in AdS/CFT correspondence that enables the calculation of the Page curve of black hole radiation, which reveals the remarkable fact that operators in island, the interior of the black hole, are in fact nonlocally dressed to the external bath, the early-time Hawking radiation. This articulated how the information in the island is encoded in the external bath.}

In this letter, we show that in the appearance of enough nontrivial matter distributions, which spontaneously break the diffeomorphism symmetry, one is able to construct such a vector field whose components are the quantum fluctuations of these matter distributions. We provide the construction to the leading order in $\kappa$ and we expect that our construction could be extended to all orders in $\kappa$. At the end, we clarify the nature of this vector field in different scenarios and their physical implications.

\section{The ADM Formalism and Diffeomorphism Constraints}
The ADM formalism \cite{Arnowitt:1962hi} in general relativity is the natural framework to study the question we discussed in the introduction, as diffeomorphism invariance is formulated as a set of algebraic constraint equations in this formalism. Diffeomorphism invariant observables are solutions of these constraint equations. In this section, we set up the stage of the calculations in this letter with a lightening review of the ADM formalism.

The ADM formalism starts with the following decomposition of metric on a (d+1)-dimensional spacetime
\begin{equation}
ds^{2}=-N^{2} dt^{2}+g_{ij}(dx^{i}+N^{i}dt)(dx^{j}+N^{j}dt)\,,\label{eq:ADM}
\end{equation}
where $N$ is called the \textit{lapse function}, the vector $N^{i}$ is called the \textit{shift vector}. The matter-coupled Einstein-Hilbert action can be written as
\begin{equation}
\begin{split}
S=&\frac{1}{16\pi G_{N}}\int dt d^{d}x N\sqrt{g}(R[g]-2\Lambda+K_{ij}K^{ij}-K^{2})\\&+S_{\text{matter}}+S_{\text{bdy}}\,,\label{eq:actionADM}
\end{split}
\end{equation}
with $\Lambda$ as the cosmological constant, $R[g]$ as the Ricci scalar of the metric $g_{ij}$, $S_{\text{matter}}$ as the action of the minimally coupled matter field and $S_{\text{bdy}}$ denoting the boundary term of the action. In the above formula, $K_{ij}$ is the extrinsic curvature of the constant-$t$ slices and it is given by
\begin{equation}
K_{ij}=\frac{1}{2N}(-\dot{g}_{ij}+D_{j}N_{i}+D_{i}N_{j})\,,\label{eq:K}
\end{equation}
where $D_{i}$ is the torsion-free and metric-compatible covariant derivative with respect to $g_{ij}$. From Equ.~(\ref{eq:actionADM}) and Equ.~(\ref{eq:K}), we can see that the following canonical momenta are zero
\begin{equation}
\Pi=\frac{1}{\sqrt{g}}\frac{\delta S}{\delta \dot{N}}=0\,,\quad \Pi_{i}=\frac{1}{\sqrt{g}}\frac{\delta S}{\delta \dot{N^{i}}}=0\,,\label{eq:primary}
\end{equation}
which are the \textit{primary constraints} \cite{dirac1938ctr}. Meanwhile, the canonical Hamiltonian of this system can be written as
\begin{equation}
H_{\text{tot}}=\int d^{d}x\sqrt{g}\Big[N\mathcal{H}+N^{i}\mathcal{H}_{i}\Big]+H_{\text{bdy}}\,,\label{eq:Htot}
\end{equation}
where
\begin{equation}
\begin{split}
  \mathcal{H}&=16\pi G_{N}\Big(\Pi_{ij}\Pi^{ij}-\frac{1}{d-1}(\Pi^{i}_{i})^{2}\Big)-\frac{1}{16\pi G_{N}}(R[g]-2\Lambda)\\&+\mathcal{H}_{\text{matter}}\,,\\
\mathcal{H}_{i}&=-2g_{ij}D_{k}\Pi^{jk}+\mathcal{H}_{i,\text{matter}}\,.\label{eq:constraints}
  \end{split}
\end{equation}
In Equ.~(\ref{eq:constraints}), we denote the Hamiltonian density of the matter field as $\mathcal{H}_{\text{matter}}$, the momentum density of the matter fields as $\mathcal{H}_{i,\text{matter}}$ and the canonical momentum of $g_{ij}$ as
\begin{equation}
\Pi^{ij}=\frac{1}{\sqrt{g}}\frac{\delta S}{\delta \dot{g}_{ij}}=-\frac{1}{16\pi G_{N}}\Big(K^{ij}-g^{ij}K\Big)\,.
\end{equation}
The primary constraints Equ.~(\ref{eq:primary}) have to be preserved under the time-evolution generated by the Hamiltonian Equ.~(\ref{eq:Htot}) and this consideration generates the \textit{secondary constraints}
\begin{equation}
\mathcal{H}=0\,,\quad \mathcal{H}_{i}=0\,.\label{eq:Hconstraints}
\end{equation}
After we promote $\mathcal{H}$ and $\mathcal{H}_{i}$ to operators $\hat{\mathcal{H}}$ and $\hat{\mathcal{H}}_{i}$, these secondary constraints constrain physical states and observables. Physical states have to be annihilated by $\hat{\mathcal{H}}$ and $\hat{\mathcal{H}}_{i}$ and physical, i.e. gauge invariant, observables have to commute with them. These constraints imposed on the states are also called the Wheeler-de Witt equations.

\section{The Quantum Rods and Clock}
We are interested in a universe with nontrivial matter distributions as this is the situation in our universe and we want to articulate how these matter distributions can be used as a reference system to define local operators. An ideal such reference system should have enough features that we can use it to define both time $x^{0}$ and locations $x^{i}$ where $i=1,\cdots,d$. Let's model the matter distributions using $(d+1)$ free scalar fields $\Phi^{\mu}(x)$ and the matter distributions are background configurations $\phi_{0}^{\mu}(x^{\mu})$, i.e. each configuration depends only on one spacetime coordinate. These background configurations and the corresponding backreacted metric $G^{(0)}_{\mu\nu}(x)$\footnote{That is $\{G^{(0)}_{\mu\nu}(x),\phi^{\mu}_{0}(x^{\mu})\}$ exactly solves the matter sourced Einstein's equations and so they satisfy the classical diffeomorphism constraints Eq.~(\ref{eq:Hconstraints}).} specifies the universe we are considering. Intuitively, we can think of these matter distributions $\phi^{\mu}_{0}(x^{\mu})$ as the rods and clock in the universe with metric $G^{(0)}_{\mu\nu}$. The final goal of us is to construct local operators in this universe which satisfies the diffeomorphism constraints.

In this section, we will study the quantum fluctuations of these rod and clock fields $\Phi^{\mu}(x)$ in the universe $\{G^{(0)}_{\mu\nu}(x),\phi^{\mu}_{0}(x^{\mu})\}$. We can write
\begin{equation}
    \Phi^{\mu}(x)=\phi_{0}^{\mu}(x^{\mu})+\phi^{\mu}(x)\,,\label{eq:linearization}
\end{equation}
where the fields $\phi^{\mu}(x)$ denotes the fluctuations. The full action of these fields in any spacetime is
\begin{equation}
    S[\Phi]=-\frac{1}{2}\int d^{d+1}x\sqrt{-G}\Big[G^{\alpha\beta}\partial_{\alpha}\Phi^{\mu}\partial^{\beta}\Phi^{\mu}+m_{\mu}^{2}(\Phi^{\mu})^{2}\Big]\,,\label{eq:clockrod}
\end{equation}
where $m_{\mu}$ is the mass of the field $\Phi^{\mu}(x)$ and the summation over $\mu$ is implicit. For the sake of convenience, we will ignore the indices $\mu$ of the field $\Phi^{\mu}$ and restore them when appropriate from now on. Under the ADM decomposition, we can write the action Equ.~(\ref{eq:clockrod}) as
\begin{equation}
   \frac{1}{2}\int d^{d+1}x\sqrt{g}N\Big[\frac{1}{N^2}\Big(\dot{\Phi}-N^{i}\partial_{i}\Phi\Big)^{2}-g^{ij}\partial_{i}\Phi\partial_{j}\Phi-m^{2}\Phi^{2}\Big]\,,
\end{equation}
from which we have the canonical momenta density
\begin{equation}
    \pi_{\Phi^{\mu}}=\frac{1}{\sqrt{g}}\frac{\delta S}{\delta\dot{\Phi}^{\mu}(x)}=\frac{1}{N}(\dot{\Phi}^{\mu}-N^{i}\partial_{i}\Phi^{\mu})\,,
\end{equation}
which obey the equal-time commutation relations
\begin{equation}
    [\hat{\pi}_{\Phi^{\mu}}(t,\vec{x}),\Phi^{\nu}(t,\vec{y})]=-i\frac{1}{\sqrt{g}}\delta^{d}(\vec{x}-\vec{y})\delta^{\nu}_{\mu}\,.\label{eq:cc}
\end{equation}
The corresponding Hamiltonian is
\begin{equation}
\begin{split}
    H[\Phi]&=\int d^{d}x\sqrt{-G} \Big[\pi(x) \dot{\Phi}(x)-\mathcal{L}\Big]\,,\\&=\int d^{d}x\sqrt{g}\Big[\frac{N}{2}\Big(\pi^{2}+g^{ij}\partial_{i}\Phi\partial_{j}\Phi+m^{2}\Phi^{2}\Big)+N^{i}\pi\partial_{i}\Phi\Big]\,,
\end{split}
\end{equation}
from which and Equ.~(\ref{eq:Htot}) we have
\begin{equation}
   \mathcal{H}[\Phi]=\frac{1}{2}\Big(\pi^{2}+g^{ij}\partial_{i}\Phi\partial_{j}\Phi+m^{2}\Phi^{2}\Big)\,,\mathcal{H}_{i}[\Phi]=\pi\partial_{i}\Phi\,.
\end{equation}
Therefore, under the linearization Equ.~(\ref{eq:linearization}) and
\begin{equation}
    G_{\mu\nu}(x)=G^{(0)}_{\mu\nu}(x)+\kappa h_{\mu\nu}(x)\,,\label{eq:gexp}
\end{equation}
we have
\begin{equation}
 \begin{split}
 &\pi_{\phi^{\mu}}=\frac{1}{N}(\dot{\phi}_{0}^{\mu}-N^{i}\partial_{i}\phi_{0}^{\mu}+\dot{\phi}^{\mu}-N^{i}\partial_{i}\phi^{\mu})\,,\\&
    [\hat{\pi}_{\phi}(t,\vec{x}),\hat{\phi}^{\nu}(t,\vec{y})]=-i\frac{1}{\sqrt{g}_{0}}\delta^{d}(\vec{x}-\vec{y})\delta^{\nu}_{\mu}\,,\label{eq:linearizationmomenta}
    \end{split}
\end{equation}
which tells us that
\begin{equation}
\begin{split}
    [\hat{\mathcal{H}}[\Phi](t,\vec{x}),\hat{\phi}^{\mu}(t,\vec{y})]&=-i\frac{1}{N\sqrt{g}_{0}}\delta^{d}(\vec{x}-\vec{y})(\dot{\phi}_{0}^{\mu}-N^{i}\partial_{i}\phi_{0}^{\mu})\,,\\&=-i\frac{1}{N\sqrt{g_{0}}}\delta^{d}(\vec{x}-\vec{y})(\delta^{\mu 0}\phi_{0}^{0\prime}-N^{i}\phi_{0}^{i\prime}\delta^{\mu}_{ i})\,,\\ [\hat{\mathcal{H}}_{i}[\Phi](t,\vec{x}),\hat{\phi}^{\mu}(t,\vec{y})]&=-i\frac{1}{N\sqrt{g_{0}}}\delta^{d}(\vec{x}-\vec{y})\partial_{i}\phi^{\mu}\,,\\&=-i\frac{1}{N\sqrt{g_{0}}}\delta^{d}(\vec{x}-\vec{y})\delta^{\mu}_{i}\phi^{i\prime}\,,\label{eq:constraintrc}
    \end{split}
\end{equation}
where we are working at the leading order in $\phi^{\mu}$ and $h_{\mu\nu}$. Last but not least, the parameter $\kappa$ in Equ.~(\ref{eq:gexp}) is chosen such that the graviton field $h_{\mu\nu}(x)$ is canonically normalized, i.e. in terms of Newton's constant $\kappa=\sqrt{32\pi G_{N}}$.

\section{Defining Local Operators Using the Quantum Rods and Clock}
In the previous section, we introduced the quantum fluctuations $\hat{\phi}^{\mu}(x)$ of the rod and clock fields in the universe $\{G^{(0)}_{\mu\nu}(x),\phi^{\mu}_{0}(x)\}$. Let's call these fluctuations the \textit{quantum rod and clock fields}. In this section, we will show that one can use these quantum rod and clock fields to construct local operators in the universe $\{G^{(0)}_{\mu\nu}(x),\phi^{\mu}_{0}(x)\}$. For the sake of simplicity, we focus on scalar operators $\hat{O}(x)$.

Intuitively, under the diffeomorphism transform
\begin{equation}
    x^{\mu}\rightarrow x^{\mu}+\kappa \epsilon^{\mu}\,,
\end{equation}
we have
\begin{equation}
    \hat{O}(x)\rightarrow\hat{O}(x)+\kappa \epsilon^{\mu}\partial_{\mu}\hat{O}(x)\,,
\end{equation}
and
\begin{equation}
    \Phi^{\mu}(x)\rightarrow\Phi^{\mu}(x)+\kappa\epsilon^{\nu}\partial_{\nu}\Phi^{\mu}(x)\,.
\end{equation}
Therefore, using Equ.~(\ref{eq:linearization}) we have that, to the leading order of the perturbation, for the quantum rod and clock fields
\begin{equation}
    \hat{\phi}^{\mu}(x)\rightarrow\hat{\phi}^{\mu}(x)+\kappa \epsilon^{\mu} \phi_{0}^{\mu\prime}(x^{\mu})\,.
\end{equation}
Let's define a more convenient quantum rod and clock fields as
\begin{equation}
    \delta\phi^{\mu}(x)\equiv\frac{\hat{\phi}^{\mu}(x)}{\phi_{0}^{\mu\prime}(x^{\mu})}\,,\label{eq:deltaphi}
\end{equation}
which under the diffeomorphism transforms as
\begin{equation}
    \delta\phi^{\mu}(x)\rightarrow\delta\phi^{\mu}(x)+\kappa\epsilon^{\mu}(x)\,.
\end{equation}
Therefore, $\delta\phi^{\mu}(x)$ is really the Goldstone fields associated with the spontaneously broken diffeomorphism symmetry and the combinations $x^{\mu}-\delta\phi^{\mu}$ are diffeomorphism invariant. Hence we expect the following operator to be diffeomorphism invariant
\begin{equation}
    \hat{O}^{\text{Phys}}(x)=\hat{O}(x-\delta\phi)\,,\label{eq:Ophys}
\end{equation}
where $x-\delta\phi$ is a shorthanded notation for $x^{\mu}-\delta\phi^{\mu}(x)$.

Let's denote the constraint operators associate with $\hat{O}(x)$ as $\hat{\mathcal{H}}[O]$ and $\hat{\mathcal{H}}_{i}[O]$ which gives,\footnote{For example, this can be worked out for a probe scalar field using canonical quantization.}
\begin{equation}
\begin{split}
    [\hat{\mathcal{H}}[O](t,\vec{x}),\hat{O}(t,\vec{y})]&=-i\frac{1}{N\sqrt{g_{0}}}\delta^{d}(\vec{x}-\vec{y})(\dot{O}-N^{i}\partial_{i}O)\,,\\ [\hat{\mathcal{H}}_{i}[O](t,\vec{x}),\hat{O}(t,\vec{y})]&=-i\frac{1}{N\sqrt{g_{0}}}\delta^{d}(\vec{x}-\vec{y})\partial_{i}O\,.
    \end{split}
\end{equation}
Using Equ.~(\ref{eq:constraintrc}), we can explicitly see that
\begin{equation}
    \begin{split}[\hat{\mathcal{H}}_{\text{matter}},\hat{O}^{\text{Phys}}(x)]=[\hat{\mathcal{H}}[O]+\hat{\mathcal{H}}[\phi],\hat{O}^{\text{Phys}}(x)]&=0\,,\\ [\hat{\mathcal{H}}_{i,\text{matter}},\hat{O}^{\text{Phys}}(x)]=[\hat{\mathcal{H}}_{i}[O]+\hat{\mathcal{H}}_{i}[\phi],\hat{O}^{\text{Phys}}(x)]&=0\,,
    \end{split}
\end{equation}
Since the above operators don't involve the graviton field, we in fact have
\begin{equation}
    [\hat{\mathcal{H}},\hat{O}^{\text{Phys}}(x)]=0\,,\quad [\hat{\mathcal{H}}_{i},\hat{O}^{\text{Phys}}(x)]=0\,,
\end{equation}
where $\mathcal{H}$ and $\mathcal{H}_{i}$ are given in Equ.~(\ref{eq:constraints}). As a result, the operators we constructed in Equ.~(\ref{eq:Ophys}) indeed satisfy the diffeomorphism constraints and qualify as local operators. Moreover, we can see that the existence of the nontrivial background configurations $\phi^{\mu}_{0}(x)$ is essential as otherwise $\phi_{0}^{\mu\prime}(x^{\mu})=0$ and the quantum rod and clock fields $\delta\phi^{\mu}(x)$ in Equ.~(\ref{eq:deltaphi}) are not well-defined.

\section{Conclusions and Discussions}
In this letter, we provide an explicit construction of local operators in a gravitational universe with nontrivial background matter distributions. Our construction follows the original insight from Einstein in the early days of general relativity that physical observables are relatively defined in a gravitational theory. We found that this is the case even quantum mechanically. More precisely, matter sources curved the spacetime and are classicalized as background configurations, which can then be thought of as a reference frame. Hence probe operators can be dressed to these background configurations, i.e. defined with respect to the corresponding reference frame. Moreover, sensible local operators can exist only if the background configurations have strong enough features to serve as a good reference frame, for example in our case $\phi_{0}^{\mu\prime}\neq0$. In our construction, we make use of the quantum fluctuations of the reference frame to compensate the change of local operators under diffeomorphism transforms. We should think of these fluctuations as components of a vector field $V^{\mu}(x)$ which however transforms as a Goldstone vector field under the diffeomorphism. The existence of this vector field is consistent with the fact that diffeomorphism symmetries are spontaneously broken by the background configurations $\phi^{\mu}_{0}(x^{\mu})$. Therefore, the construction in this paper is in the same spirit as the constructions in \cite{Geng:2023zhq,Geng:2024tba1,Geng:2024tba2} in a specific theory of massive gravity. In that case, the Goldstone vector field $V^{\mu}(x)$ is a composite operator \cite{Geng:2023ynk,Geng:2024tba1,Geng:2024tba2}. Moreover, correlators of the physical operators like $\hat{O}^{\text{Phys}}(x)$ can be straightforwardly computed in the unitary gauge, i.e. $V^{\mu}(x)=0$, which reduces the computation to quantum field theory in the fixed background.\footnote{A caveat is that the state under which we are computing the correlator also has to satisfy the diffeomorphism constraints, i.e. it must obey the Wheeler-de Witt equations.} It deserves to be emphasized that the construction in this letter works only to the first nontrivial order in the linearization Equ.~(\ref{eq:linearization}) and Equ.~(\ref{eq:gexp}), where graviton field and the matter fields are decoupled and based on the above discussion we expect the construction can be extended to all orders in the expansion which is in fact counted by the powers of $\kappa$, i.e. the Newton's constant. We leave this construction for future work.\footnote{A good starting point is low dimensional models of gravity coupled with matter \cite{Wu:2023uyb}.}

We hope that the current letter clarifies some open questions in the recent literature \cite{Chandrasekaran:2022cip,Folkestad:2023cze,Witten:2023qsv,Witten:2023xze,Jensen:2023yxy,Chen:2024rpx,Kudler-Flam:2024psh} regarding subregion algebras in a gravitational universe (see also \cite{RevModPhys.33.510,PhysRev.124.274,CRovelli_1991,CRovelli_19912,Nicolis:2015sra,Bose:2017nin,Piazza:2021ojr,Piazza:2022amf,Carrozza:2022xut,Goeller:2022rsx,Hoehn:2023axh,AliAhmad:2023etg,Bamonti:2023tjb,Torrieri:2024ivy,DeVuyst:2024pop,AliAhmad:2024eun,AliAhmad:2024saq,AliAhmad:2024vdw,AliAhmad:2024wja,Grassi:2024vkb} for relevant discussions). For example, to define subregion algebras in gravity, one doesn't need the existence of an ``observer" with a non-physical Hamiltonian and that ``observer" can be taken as a simplified model of a nontrivial background configuration which defines the subregion. Such background configurations universally exist in cosmological scenarios, for example during the inflation the classical inflaton field is time-dependent and its quantum fluctuation can be used as a clock field according to our construction. Interestingly, this consideration is in fact implicit in the construction of effective actions for perturbations in a inflationary universe \cite{Cheung:2007st,Gubitosi:2012hu,Piazza:2013coa}.

\section*{Acknowledgements}
We would like to thank Steve Giddings, Daniel Harlow, $\mathring{\text{A}}$smund Folkstad, Jonah Kudler-Flam, Juan Maldacena, Henry Maxfield, Massimo Porrati, Lisa Randall, Edward Witten and Jiuci Xu for discussions. We would like to thank Daniel Jafferis, Pushkal Shrivastava and Neeraj Tata for relevant collaborations. HG is supported by the Gravity, Spacetime, and Particle Physics (GRASP) Initiative from Harvard University.
\bibliographystyle{apsrev4-1}
\bibliography{main}
\end{document}